\documentclass[aps,prd,twocolumn,superscriptaddress,amsmath,amssymb,nofootinbib]{revtex4-2}
\usepackage{graphicx} 
\usepackage{epstopdf} 
\usepackage{dcolumn} 
\usepackage{xcolor} 
\usepackage{amsmath,amssymb} 
\usepackage{hyperref} 
\usepackage[utf8]{inputenc} 
\usepackage[english]{babel} 
\usepackage{ifthen} 
\usepackage{orcidlink} 
\usepackage{physics}

\graphicspath{{figures/}} 

\newboolean{articletitles}
\setboolean{articletitles}{true} 

\input{belle2-symbols}

\newcommand{\Asymm}{\ensuremath{\mathcal{A}}\xspace}
\newcommand{\ACP}{\ensuremath{\Asymm_{\CP}}\xspace}
\newcommand{\raw}{{\rm raw}}
\newcommand{\Araw}{\ensuremath{\Asymm_{\raw}}\xspace}
\newcommand{\Aprod}{\ensuremath{\Asymm_{\rm prod}}\xspace}
\newcommand{\Adet}{\ensuremath{\Asymm_{\varepsilon}}\xspace}
\newcommand{\tagged}{{\rm tagged}}
\newcommand{\untagged}{{\rm untagged}}
\newcommand{\Arawt}{\ensuremath{\Araw^\tagged}\xspace}
\newcommand{\Arawu}{\ensuremath{\Araw^\untagged}\xspace}

\newcommand{\Adetpitag}{\ensuremath{\Adet^{\pitag}}\xspace}
\newcommand{\splt}[1][]{\ensuremath{{_{S}\mathcal{P}\mathit{lot#1}}}\xspace}

\newcommand{\sig}{{\rm sig}}
\newcommand{\csig}{{\rm Csig}}
\newcommand{\misreco}{{\rm Msig}}
\newcommand{\rpi}{{\mathrm{rand}}}

\newcommand{\pkg}{{K\pi\piz}}
\newcommand{\npk}{{\rm comb}}

\newcommand{\pitag}{\ensuremath{\pi_\mathrm{tag}}\xspace}

\newcommand{\pipipiz}{\ensuremath{\pi\pi\piz}\xspace}
\newcommand{\DzTopipipiz}{\texorpdfstring{\ensuremath{\Dz\to\pip\pim\piz}}{D0 -> pi+ pi- pi0}\xspace}
\newcommand{\DzbTopipipiz}{\ensuremath{\Dzb\to\pip\pim\piz}\xspace}

\newcommand{\DstarpToDzpi}{\ensuremath{\Dstarp\to\Dz\pip}\xspace}

\newcommand{\DzToKpipiz}{\ensuremath{\Dz\to\Km\pip\piz}\xspace}
\newcommand{\DzToKpiRS}{\texorpdfstring{\ensuremath{\Dz\to\Km\pip}}{D0 -> K- pi+}\xspace}
\newcommand{\DzbToKpiRS}{\ensuremath{\Dzb\to\Kp\pim}\xspace}
\newcommand{\DzToKpiWS}{\ensuremath{\Dz\to\Kp\pim}\xspace}

\newcommand{\DzToKspiz}{\ensuremath{\Dz\to\KS\piz}\xspace}

\newcommand{\thetacm}{\ensuremath{\theta_{\rm CM}}\xspace}
\newcommand{\costhetacm}{\ensuremath{\cos\thetacm}\xspace}
\newcommand{\Dzcosthetacm}{\ensuremath{\cos\smash{\thetacm^{\Dz}}}\xspace}
\newcommand{\Dstarpcosthetacm}{\ensuremath{\cos\smash{\thetacm^{\Dstarp}}}\xspace}
\newcommand{\dm}{\ensuremath{{\Delta M}}\xspace}

\RequirePackage{amsopn}

\newcommand{\tesla}{\ensuremath{{\rm \,T}}\xspace}

\def\mubar{\kern 0.2em\overline{\kern -0.2em\mu}{}\xspace}

\newcommand{\bii}{\texorpdfstring{Belle~II}{Belle II}\xspace}
\newcommand{\skb}{SuperKEKB\xspace}
\newcommand{\wrt}{with respect to\xspace}

\newcommand{\lumi}{\ensuremath{{428\invfb}}\xspace}

\RequirePackage{adjustbox}

\begin{document}

\vspace*{-3\baselineskip}
\resizebox{!}{2cm}{\includegraphics{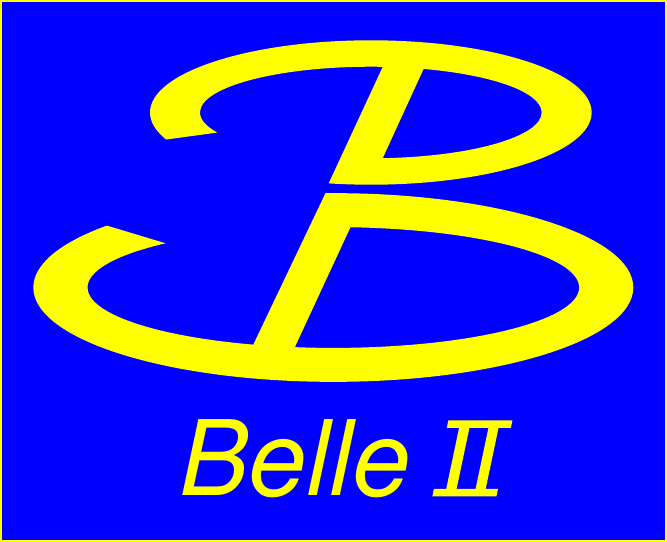}}

\title{Measurement of the \CP asymmetry in \DzTopipipiz decays at \bii}
  \author{M.~Abumusabh\,\orcidlink{0009-0004-1031-5425}} 
  \author{I.~Adachi\,\orcidlink{0000-0003-2287-0173}} 
  \author{L.~Aggarwal\,\orcidlink{0000-0002-0909-7537}} 
  \author{H.~Ahmed\,\orcidlink{0000-0003-3976-7498}} 
  \author{Y.~Ahn\,\orcidlink{0000-0001-6820-0576}} 
  \author{H.~Aihara\,\orcidlink{0000-0002-1907-5964}} 
  \author{N.~Akopov\,\orcidlink{0000-0002-4425-2096}} 
  \author{S.~Alghamdi\,\orcidlink{0000-0001-7609-112X}} 
  \author{M.~Alhakami\,\orcidlink{0000-0002-2234-8628}} 
  \author{A.~Aloisio\,\orcidlink{0000-0002-3883-6693}} 
% ...
\collaboration{The Belle II Collaboration}

\begin{abstract}
We measure the time- and phase-space-integrated \CP asymmetry \ACP in \DzTopipipiz decays reconstructed in \(\epem\to\ccbar\) events collected by the \bii experiment from 2019 to 2022.
This sample corresponds to an integrated luminosity of \lumi.
We require \Dz mesons to be produced in \DstarpToDzpi decays to determine their flavor at production.
Control samples of \DzToKpiRS decays are used to correct for reconstruction-induced asymmetries.
The result, \(\ACP(\DzTopipipiz)=(0.29\pm0.27\pm0.13)\%\), where the first uncertainty is statistical and the second systematic, is the most precise result to date and is consistent with \CP conservation.

\end{abstract}
\preprint{\bii preprint: 2025-018}
\preprint{KEK preprint: 2025-17}

\maketitle

\section{Introduction\label{sec:intro}}
Searches for charge-parity (\CP) violation in the charm sector provide a unique opportunity to explore possible physics beyond the standard model (SM) and are complementary to measurements in the strange and beauty sectors, especially for models of new physics where up-type quarks have a special role.
\CP violation is predicted to be very small in charm transitions because the contribution from the third generation of quarks is highly suppressed \cite{CharmTwoGenerations}.
The largest SM asymmetries, which are expected to occur in singly-Cabibbo-suppressed channels, are predicted to be of order \(10^{-4}\)--\(10^{-3}\) \cite{Golden:1989qx,Grossman:2006jg}.

Experimental sensitivity has reached the level of these predictions only in recent years:
the first and only observation of \CP violation in the charm sector was performed in 2019 by the LHCb Collaboration \cite{onlyCharmCPV,onlyCharmCPV2}.
Its origin is not yet understood, with both new physics and unaccounted-for nonperturbative QCD contributions being possible explanations \cite{Chala:2019fdb,Dery:2019ysp,Gavrilova:2023fzy}.
In this context, searching for \CP violation in additional channels and improving the precision of previous measurements are essential.

In this paper, we report a measurement of the time- and phase-space-integrated \CP asymmetry in \DzTopipipiz decays using \(\epem\to\ccbar\) events collected by \bii between 2019 and 2022.
This dataset corresponds to an integrated luminosity of \lumi.
The time-integrated \CP asymmetry is defined as
\begin{multline}
    \ACP(\DzTopipipiz) = \\
    \frac{\Gamma(\DzTopipipiz)-\Gamma(\DzbTopipipiz)}{\Gamma(\DzTopipipiz)+\Gamma(\DzbTopipipiz)},
\end{multline}
where \(\Gamma\) indicates the decay-time-integrated decay rates, which include \Dz--\Dzb mixing effects.
The most precise measurement of this observable to date, \((0.31\pm0.41\pm0.17)\%\) where the first uncertainty is statistical and the second systematic, was performed by the \babar Collaboration using about \(82\times10^3\) signal candidates reconstructed in a \(385\invfb\) dataset \cite{babarACP}.
Other measurements for this channel include an \ACP measurement by \belle \cite{belleACP}, an unbinned statistical test of the Dalitz plot distribution symmetry by \lhcb \cite{LHCbEnergyTest}, and a time-dependent \CP violation search by \lhcb \cite{LHCbDeltaY}.
All are compatible with \CP symmetry.

The flavor of the neutral \D meson is identified (or ``tagged'') by requiring that the meson originates from a \DstarpToDzpi decay.
(Throughout this paper, \CP-conjugate decays are implied unless stated otherwise.)
In this case, the charge of the \pip identifies the flavor of the \Dz at production.
We refer to this low-momentum charged pion as the ``tag pion'', \pitag.

To determine \ACP, we measure the raw asymmetry between the number of reconstructed decays of the two flavors:
\begin{equation}\label{eq:araw}
    \Araw^{\pipipiz}
    = \frac{N(\DzTopipipiz)-N(\DzbTopipipiz)}{N(\DzTopipipiz)+N(\DzbTopipipiz)}.
\end{equation}
This asymmetry has contributions from several sources.
Given the small magnitude of the contributions (a few percent at most) we can approximate it as a sum:
\begin{equation}
    \Araw^{\pipipiz} \simeq \ACP + \Aprod + \Adet^{\pipipiz} + \Adetpitag.
\end{equation}
The \CP asymmetry \ACP is the observable of interest;
the production asymmetry \Aprod arises from the forward-backward asymmetric \Dstarp production in \(\epem\to\ccbar\) processes, and
the terms \(\Adet^{\pipipiz}\) and \Adetpitag result from asymmetric efficiencies in the reconstruction of the \Dz meson and tag pion, respectively.

To measure \Adetpitag, we use two control samples of \DzToKpiRS decays with and without reconstruction of the \DstarpToDzpi decay:
we refer to these as the tagged and untagged samples, respectively.
Being dominated by a Cabibbo-favored \(\cquark\to\squark\) transition, this decay mode is both abundant and self-tagging.
In addition, it can be reconstructed with high purity and efficiency due to the presence of two charged particles and no neutral particle in the final state.
The raw asymmetries of these samples are
\begin{subequations}
    \begin{align}
        \Arawt &\simeq \Aprod + \Adet^{K\pi} + \Adetpitag \\
        \Arawu &\simeq \Aprod + \Adet^{K\pi},
    \end{align}
\end{subequations}
where we have neglected a possible \CP asymmetry, which is expected to be negligible for a Cabibbo-favored decay at our current level of sensitivity \cite{Golden:1989qx,Grossman:2006jg}.
From these we compute
\begin{equation}\label{eq:AtagDef}
    \Adetpitag = \Arawt - \Arawu
\end{equation}
which we then subtract from \(\Araw^{\pipipiz}\).

Given that the \Dz final state is self-conjugate, and that final-state particles have relatively high momenta on average, we expect the \Dz reconstruction asymmetry \(\Adet^{\pipipiz}\) to be negligible.
We confirm this in simulation and assign a systematic uncertainty for this choice (see \autoref{sec:syst}).

The production asymmetry is caused by \g--\Z interference and higher-order QED effects in the \(\epem\to\ccbar\) process \cite{Afb}.
It is an odd function of the cosine of the polar angle \costhetacm of the charm quark momentum in the collision c.m. frame
and reaches a maximum of \(\order(1\%)\).
Since the strong interaction responsible for hadronization and \DstarpToDzpi decays is \CP conserving, \Aprod is also an odd function of the \costhetacm of the \Dstarp and \Dz mesons, \Dstarpcosthetacm and \Dzcosthetacm, which we measure.
Because the reconstruction efficiency is not symmetric in \costhetacm, the production asymmetry does not cancel when integrated over all reconstructed \Dstar mesons.
To cancel \Aprod, we divide our samples into eight bins of \Dzcosthetacm of the \Dz meson.
The bins are chosen to be symmetric around \(\Dzcosthetacm=0\) and small enough such that the efficiency is approximately constant within each bin.
We measure the \DzTopipipiz raw asymmetries separately in each oppositely signed \Dzcosthetacm bin \(\pm i\) (with \(i=1,\dots,4\)), correct them for the tag pion asymmetries computed in the same bin, and determine \ACP from the arithmetic average of positive and negative bins to cancel the odd contribution from \Aprod,
\begin{equation}\label{eq:AfinalDef}
    \ACP^i = \frac{\Asymm_{+i}+\Asymm_{-i}}{2} \,,
\end{equation}
where \(\Asymm_{\pm i} = \Asymm_{\raw,\pm i}^{\pipipiz} - \Asymm^{\pitag}_{\varepsilon,\pm i}\).
We then obtain \ACP as the average of the \(\ACP^i\) values.

We ensure that the detection and production asymmetries of signal and control samples are the same by employing the same selection criteria for variables that the asymmetries may depend on,
and by using per-candidate weights to equalize the distributions of kinematic variables that the asymmetries are sensitive to.
To avoid potential bias, the measured values of the raw \DzTopipipiz asymmetries in each \Dzcosthetacm bin were shifted by an undisclosed offset until the entire measurement procedure was finalized, and all systematic uncertainties were computed.

The remainder of this paper is organized as follows.
\autoref{sec:b2} briefly describes the \bii detector and the simulation samples used.
\autoref{sec:selections} describes the reconstruction process and the selection criteria used for the signal and the two control samples.
\autoref{sec:weights} describes the weighting procedure that ensures the correct estimation of reconstruction asymmetries.
\autoref{sec:fits} describes the fit procedures used to measure the raw signal asymmetries.
\autoref{sec:syst} discusses the sources of systematic uncertainty.
Finally, \autoref{sec:results} presents our results.

\section{\bii detector and samples\label{sec:b2}}
The \bii detector \cite{b2tdr,b2physbook}, located at the beam interaction region (IR) of the \skb asymmetric-energy \epem collider \cite{skb}, is a large-solid-angle spectrometer.
It has a cylindrical geometry and consists of (from inner to outer radius):
a silicon vertex detector made of two layers of pixel sensors, plus four layers of double-sided strip sensors \cite{SVD};
a 56-layer drift chamber;
a barrel time-of-propagation detector \cite{TOP}, and a forward-end-cap aerogel ring-imaging Čerenkov detector; and
an electromagnetic calorimeter made of CsI(Tl) crystals.
These subdetectors operate within a \(1.5\tesla\) magnetic field produced by a superconducting solenoid.
An iron flux-return yoke outside the solenoid is instrumented with resistive plate chambers and plastic scintillator tiles to detect muons and \KL mesons.
For the dataset used in this analysis, only part of the second layer of pixel detector was installed, corresponding to one-sixth of the azimuthal angle.
The longitudinal \(z\) axis of the laboratory frame is defined as the central axis of the solenoid, with the positive direction given by the direction of the electron beam.

In this work we use the data sample collected from the beginning of data taking in 2019 until 2022, which corresponds to an integrated luminosity of \lumi.
This sample includes collisions with energy on the \FourS resonance (about \(365\invfb\)), below the \BBbar pair production threshold (about \(44\invfb\)), and at various values around the \FiveS resonance (about \(19\invfb\)).
The variation in \(\epem\to\ccbar\) cross section over this range of collision energies is less than \(5\%\).

We use simulated Monte Carlo (MC) samples to identify sources of background, optimize selection criteria, develop the kinematic weighting procedure, determine fit models, validate the analysis procedure, and determine some of the systematic uncertainties.
These samples correspond to four times the data integrated luminosity.
We use \evtgen \cite{evtgen} interfaced to \pythia \cite{pythia} and \kkmc \cite{kkmc} to generate \(\epem\to\FourS\) and \(\epem\to\qqbar\) events (where \quark is a \uquark, \dquark, \cquark, or \squark quark), and to simulate particle decays;
\tauola \cite{tauola} to generate \(\epem\to\tautau\) events;
\photos \cite{photos1,photos2} to simulate final-state radiation; and
\geant \cite{geant4} to simulate the interaction of particles with the detector material.
We take beam-induced backgrounds from delayed Bhabha trigger data, and overlay them on simulated events.

To simulate \DzTopipipiz decays, we employ a Dalitz distribution model based on an amplitude analysis performed by the \babar Collaboration \cite{BaBarAmplitude}.
Additionally, we scale signal and background components to better match the data distributions.
Data-MC differences for signal and charm backgrounds are of order \(10\%\) and arise mainly from incorrect fragmentation modeling in the event generator.
For backgrounds due to random combinations of final-state particles, discrepancies are of order \(1\%\).
The rescaling effectively reduces the largest differences to a few percent.
After the rescaling, the simulated distributions of the invariant masses and the Dalitz plot variables agree with those found on data.
We use the \bii analysis software framework \cite{basf2,basf2-zenodo} to process the data and MC samples.

\section{Candidate reconstruction and selection criteria\label{sec:selections}}
We consider only events with at least three tracks (trajectories of charged particles as reconstructed in the drift chamber and vertex detector) that originate from the IR and have transverse momentum \pt greater than \(200\mevc\).
These events must also be inconsistent with Bhabha scattering.
\autoref{ss:selections:sig} describes the reconstruction of the signal sample, while \autoref{ss:selections:Kpi} describes the reconstruction of the control samples.

\subsection{\DzTopipipiz signal sample\label{ss:selections:sig}}
We reconstruct photon candidates from localized energy deposits (clusters) in the calorimeter that are not geometrically matched to a track.
These clusters must have a polar angle within the acceptance of the drift chamber (\(17\degrees<\theta<150\degrees\)) to ensure they are not produced by an undetected charged particle.

To suppress beam background clusters, we require that the clusters
include energy deposits from at least two crystals,
have energies greater than \(100\mev\), and
have crystal hit times within \(200\ns\) of the measured \epem collision time;
the difference between the hit times and the collision time must also be less than twice its uncertainty.
Since the signal distribution is non-Gaussian, the latter criterion selects \(99\%\) of the correctly reconstructed photons on simulation.

Furthermore, to suppress hadronic clusters, we use selections based on the distribution of the cluster energy in the plane orthogonal to the photon momentum.
We use the ratios \(E_1/E_9\) and \(E_9/E_{21}\),
where \(E_1\), is the energy deposit in the central crystal of the cluster (the one with the highest energy),
\(E_9\) is the energy deposit in the \(3\times3\) array of crystals around the cluster center,
and \(E_{21}\) is the energy deposit in the \(5\times5\) array of crystals around the cluster center with the four corners removed.
We also use
a boosted decision tree classifier that exploits the Zernike moments of the energy deposit distribution among the crystals \cite{zernike,fastbdt}.

We combine pairs of photon candidates to form \(\piz\to\g\g\) candidates.
We require that the invariant mass of the two photons be between \(116\) and \(150\mevcc\)
and that the \piz momentum be greater than \(0.5\gevc\).
This results in a mass resolution of about \(6.5\mevcc\).

We require that charged particle tracks originate from within the IR:
the point of closest approach of a track to the \(z\) axis must be less than \(1\cm\) away from the beam interaction point along the \(z\) axis and less than \(0.5\cm\) away in the plane transverse to the \(z\) axis.
Furthermore, we require the polar angles of the track momenta to be within the acceptance of the drift chamber for consistency.

We combine pairs of opposite-charge pion candidates with \piz candidates to form \Dz candidates.
To suppress background with charged kaons misidentified as pions (\eg, from \DzToKpipiz decays), we require the charged particles' neural-network-based particle identification (NNPID) output to be greater than \(0.2\).
The NNPID is performed using information from all subdetectors except the pixel detector \cite{pidnn}.
This selection is over \(95\%\) efficient and has a kaon-to-pion misidentification rate below \(15\%\).

To reject random combinations of pions, we require that the \Dz candidate invariant mass \(M\) be between \(1.785\) and \(1.95\gevcc\).
The mass resolution is about \(20\mevcc\).
Additionally, to reject \DzToKspiz decays, the invariant mass of the charged pion pair must not be between \(470\) and \(530\mevcc\).

We combine \Dz candidates with charged pion candidates to form \DstarpToDzpi candidates.
These charged pions are required to have a NNPID output greater than \(0.1\);
this selection is more than \(97\%\) efficient, with a kaon-to-pion misidentification rate below \(20\%\).
To suppress random combinations of particles, we require that the difference between the invariant masses of the \Dstarp and \Dz candidates, \dm, be between \(140\) and \(150\mevcc\).

We perform a vertex fit that exploits kinematic and geometric information from the whole decay chain, with the constraint that the \Dstarp vertex must be in the IR \cite{treefit}.
We keep only candidates for which the fit converges.
In events with multiple candidates (about \(10\%\) of the total), we only retain the candidate with the lowest \chisq value (highest fit probability) resulting from the vertex fit.
When a correctly reconstructed signal candidate is present in a multicandidate event, this criterion selects it with \(61\%\) efficiency.

To suppress \Dstarp candidates produced in the decay of a \B meson, which are subject to different production asymmetries, we require the c.m.\ momentum of the \Dstarp candidate to be greater than \(1.06\) times the maximum kinematically allowed value for a \Dstarp meson arising from a \B decay.
This factor accounts for the smearing of the c.m.\ momentum distribution due to the finite detector resolution, and is determined from simulation.
The maximum momentum depends on the collision energy and corresponds to about \(2.45\gevc\) for the majority of the dataset.
This selection rejects about \(15\%\) of the reconstructed signal \Dstarp in \(\epem\to\ccbar\) events, over \(99\%\) of the \Dstarp produced in the decay of a \B meson, and about \(57\%\) of the combinatorial background.

Finally, we remove \Dz candidates with \(\abs*{\Dzcosthetacm}\) larger than \(0.7\) due to the presence of large backgrounds that are difficult to model.
The very forward and backward regions of the detector are affected by a much larger combinatorial background, mainly due to beam-induced showers reaching the calorimeter end caps.
Also, the resolution on track parameters is worse at extreme polar angles, and this makes it more difficult to model the background from \DzToKpipiz (see \autoref{ss:fits:sig}).

The signal selection efficiency calculated from MC samples is \(9.0\%\).
Based on this, the expected signal yield in the data sample, about \(270\times10^3\) candidates, is more than three times that of the \babar measurement \cite{babarACP}, thanks to the relaxed selection criteria and better detector performance.

\subsection{\DzToKpiRS control samples\label{ss:selections:Kpi}}
We select charged pion and kaon candidates with the same polar angle and impact parameter criteria as those used to select pions in the signal sample.

We form \Dz candidates by combining one kaon candidate with one opposite-charge pion candidate.
For these kaons and pions, we require a  NNPID output greater than \(0.8\) to suppress \Dz candidates with the \kaon and \(\pi\) each misidentified as the other, to which we assign the incorrect flavor.
This selection is over \(80\%\) efficient and has a misidentification rate below \(2\%\).
Additionally, we require the \kaon and \(\pi\) transverse momenta to be greater than \(150\mevc\), the \(\cos\theta\) of the \(\pi\) to be greater than \(-0.6\), and the \(\cos\theta\) of the \kaon to be greater than \(-0.75\).
These criteria exclude regions of the phase space where the kinematic weighting procedure (described in \autoref{sec:weights}) has poor performance.
The lower performance arises from lower average track momentum due to the forward boost of the collision c.m., and the lack of particle identification detectors for \(\theta>120\degrees\).

We require the \Dz candidate invariant mass to be between \(1.8\) and \(1.92\gevcc\)
and the \Dz c.m.\ momentum to be greater than \(1.06\) times the maximum kinematically allowed value for a \Dz meson produced in the decay of a \B meson.

We use the same selection criteria for \Dz candidates for both the tagged and untagged control samples.
This ensures we have the same reconstruction-induced asymmetries in both samples.

To select the untagged sample, we perform a vertex-and-kinematic fit \cite{treefit} on the \Dz candidates and keep only candidates with a fit probability greater than \(10^{-3}\).
In events with multiple candidates (about \(10\%\)), we select one candidate randomly.

To form the tagged sample, we begin with the untagged \Dz candidates obtained before applying the vertex fit selection.
We combine these \Dz candidates with pion candidates that have the expected charge sign to form \DstarpToDzpi candidates.
The tag pion selection is the same as that used for the signal sample.
This is essential in order to have the same reconstruction asymmetry.
We then perform a vertex-and-kinematic fit as for the untagged sample and keep only candidates with a fit probability greater than \(10^{-3}\).
In events with multiple candidates (about \(8\%\)), we select one candidate randomly.

\section{Kinematic weighting of the control samples\label{sec:weights}}
Using the same selection criteria is not sufficient to ensure that the detection and production asymmetries are the same for the different samples.
The reconstruction asymmetries also depend mainly on the momenta and polar angles of the particles.
Also, the production asymmetry depends on the distribution of the cosine of the \Dz polar angle in the collision c.m.\ frame.
Even with the same selection criteria applied to the individual final-state particles, these distributions differ due to the correlations among the kinematic variables of different particles in the decay chain.

To remove these differences, we apply per-candidate weights to the tagged and untagged samples in two steps.
First, we apply weights to the tagged sample so that its tag pion \((\pt,\cos\theta)\) distribution matches that of the signal sample.
This is done to match \Adetpitag of the two samples.
Then, we apply weights to the untagged sample so that its distribution of the kaon and pion \((\pt,\cos\theta)\), and \Dzcosthetacm, matches that of the weighted tagged control sample.
This is done to match \(\Adet^{K\pi}\) and \Aprod of the two samples.

The weights are computed using background-subtracted signal and control sample distributions.
The background subtraction is performed using the \splt technique \cite{splot} and exploiting the signal and background fit models described in \autoref{sec:fits}.
The first weighting is performed directly in two dimensions:
the weights are determined from the bin-by-bin ratio of two-dimensional \((p_{\rm T}^{\pitag},\cos\theta^{\pitag})\) histograms, with Gaussian smoothing used to reduce fluctuations.
The second weighting is performed by iteratively updating the weights using one-dimensional histogram ratios of one variable at a time.
These five one-dimensional updates of the weights are iterated several times, until the change in weights is smaller than the associated statistical uncertainty.
This is done because the sample size is insufficient for five-dimensional  \((p_{\rm T}^{\pi},\cos\theta^{\pi},p_{\rm T}^{K},\cos\theta^{K},\Dzcosthetacm)\) histogram weighting.

\begin{figure*}[htbp]
    \adjincludegraphics[width=0.325\linewidth,page=1,clip,trim={{0.375\width} 0 {0.5\width} 0}]{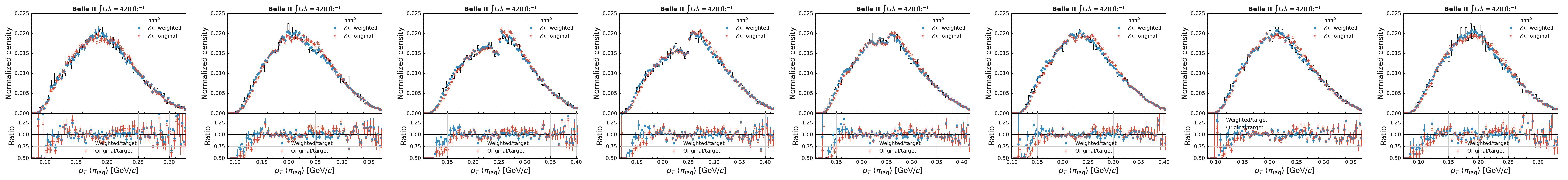}
    \adjincludegraphics[width=0.325\linewidth,page=2,clip,trim={{0.375\width} 0 {0.5\width} 0}]{data-sample}\\
    \adjincludegraphics[width=0.325\linewidth,page=3,clip,trim={{0.375\width} 0 {0.5\width} 0}]{data-sample}
    \adjincludegraphics[width=0.325\linewidth,page=4,clip,trim={{0.375\width} 0 {0.5\width} 0}]{data-sample}\\
    \adjincludegraphics[width=0.325\linewidth,page=5,clip,trim={{0.375\width} 0 {0.5\width} 0}]{data-sample}
    \adjincludegraphics[width=0.325\linewidth,page=6,clip,trim={{0.375\width} 0 {0.5\width} 0}]{data-sample}
    \adjincludegraphics[width=0.325\linewidth,page=7,clip,trim={{0.375\width} 0 {0.5\width} 0}]{data-sample}
    \caption{%
        Background-subtracted distributions of the variables used for the kinematic weighting for \(\Dzcosthetacm\in[-0.208,0)\).
        Top row: \pt and \(\cos\theta\) of the tag pion in the signal and tagged samples.
        Middle row: \pt and \(\cos\theta\) of the kaon in the untagged and tagged samples.
        Bottom row: \pt and \(\cos\theta\) of the pion, and \Dzcosthetacm of the \Dz meson, in the untagged and tagged samples.
        The black lines show the distribution for the sample that we aim to match, \ie, the signal sample in the top row and the tagged sample in the middle and bottom rows.
        The circles show the distribution for the other sample before (red empty circles) and after (blue filled circles) applying the weights.
        The bottom plots of each panel show the ratios between circles and black line.%
        \label{fig:weights}%
    }
\end{figure*}

All these steps are performed independently for each \Dzcosthetacm bin.
The effect of the weighting is shown in \autoref{fig:weights} for one \Dzcosthetacm bin.

\section{Fit model and asymmetry determination\label{sec:fits}}
Asymmetries are extracted through unbinned extended maximum-likelihood fits to the \Dz candidate invariant mass \(M\), and the mass difference between \Dstarp and \Dz candidates \dm.
Asymmetries are obtained separately in eight bins of \Dzcosthetacm to correct for the production asymmetry, as described in \autoref{sec:intro}.
The eight bins of \costhetacm are \(\pm[0,0.208)\), \(\pm[0.208,0.411)\), \(\pm[0.411,0.599)\), and \(\pm[0.599,0.7)\).

\subsection{\DzTopipipiz signal sample\label{ss:fits:sig}}
For the signal sample, we use the two-dimensional distributions of \(M\) and \dm to discriminate four components:
correctly tagged \DzTopipipiz decays;
correctly reconstructed \DzTopipipiz decays paired with an unrelated tag pion;
\DzToKpipiz decays with the kaon misidentified as a pion, and
candidates made from random combinations of final-state particles (referred to as combinatorial background).
The probability density function (PDF) of each component is taken to factorize into the product of two one-dimensional PDFs:
\begin{equation}
    p_j = p_j(M,\dm) = p_{j,M}(M)\cdot p_{j,\dm}(\dm)
\end{equation}
where \(j\) runs over the four components.
This assumption neglects small correlations between \(M\) and \dm for some backgrounds, and we assign a systematic uncertainty (see \autoref{sec:syst}) to account for this.

Correctly tagged signal peaks in both variables.
There are two subcomponents:
one where the \DzTopipipiz decay is correctly reconstructed
and another where one of the final-state particles in a \DzTopipipiz decay is misreconstructed (\eg, one of the charged pions decayed in the detector volume and was partially reconstructed from its decay products).
In both cases, since the correct tag pion is used, the correct flavor is assigned.
For the first subcomponent, the \(M\) and \dm distributions are each described by a Johnson's \(S_U\) PDF \cite{johnson},
\begin{equation}
    p(x|\mu,\lambda,\gamma,\delta) \propto \frac{\exp(-\frac{1}{2}\qty(\gamma+\delta\sinh^{-1}\qty(\frac{x-\mu}{\lambda}))^2)}{\sqrt{1+\qty(\frac{x-\mu}{\lambda})^2}}\,,
\end{equation}
where \(\mu\) is a location parameter, \(\lambda\) is a width parameter, and \(\gamma\) and \(\delta\) are shape parameters.
For the second, we use the same \dm PDF as the signal but with a different width parameter \(\lambda\) to account for its worse resolution,
and a second-order Chebyshev polynomial for \(M\).

The random-tag-pion component shares the correctly reconstructed signal \(M\) PDF but uses a thresholdlike function for \dm:
\begin{equation}\label{eq:pdf:thr}
    p_\rpi(\dm) \propto \sqrt{\dm-x_0} + p_0 + p_1(\dm-x_0),
\end{equation}
where \(x_0\) is fixed to the nominal charged pion mass \cite{pdg}.
The \DzToKpipiz background is modeled by a double-sided asymmetric Crystal Ball PDF \cite{CB1,CB2} for \dm, and a power law for \(M\).
The combinatorial component shares the same \dm PDF as the random-tag-pion component, and \(M\) is modeled by a second-order Chebyshev polynomial.

The total extended fit function is given by
\begin{align}
    &p_{\Dz,\Dzb}(M,\dm)=\nonumber\\
    &\qquad n_\sig\frac{1\pm\Asymm_{\raw,\sig}}{2}\left[(1-f_\misreco)p_\csig + f_\misreco\cdot p_\misreco\right]\nonumber\\
    &\qquad +n_\sig\cdot f_\rpi \frac{1\pm\Asymm_{\raw,\rpi}}{2}p_\rpi\nonumber\\
    &\qquad +n_\pkg \frac{1\pm\Asymm_{\raw,\pkg}}{2}p_\pkg\nonumber\\
    &\qquad +n_\npk \frac{1\pm\Asymm_{\raw,\npk}}{2}p_\npk\,,
\end{align}
where the plus sign is for \Dz mesons and the minus for \Dzb mesons.
We use \(n_j\) for the yield of the component \(j\) (including both \Dz and \Dzb mesons),
while \(f_j\) indicates the fraction of (sub)component \(j\) \wrt the correctly tagged signal.
The subscript ``\(\sig\)'' refers to the signal component,
``\(\csig\)'' refers to the correctly reconstructed signal subcomponent,
``\(\misreco\)'' refers to the subcomponent where a signal decay is reconstructed with an incorrect \Dz pion,
``\(\rpi\)'' refers to the random-tag-pion component,
``\(\pkg\)'' refers to the \DzToKpipiz background component,
and ``\(\npk\)'' refers to the combinatorial component.

The asymmetry fit is performed simultaneously in all \Dzcosthetacm bins and for both flavors.
Some parameters are shared among \Dzcosthetacm bins to improve the accuracy of their determination.
The values of most shape parameters are fixed to the values found when fitting the MC sample.
The width in each \costhetacm bin is fixed to the value obtained from MC simulation, but a global scale factor is floated in the fit to account for data-simulation differences.
Location parameters are also floated.
Finally, the fractions of random-tag-pion and misreconstructed signal components are fixed to their MC values, as they are too small to determine in data: \(f_\rpi\simeq2\%\) and \(f_\misreco\simeq5\%\).
The same applies to the asymmetry of the random-tag-pion component.

\begin{figure*}[htbp]
    \includegraphics[width=0.49\linewidth]{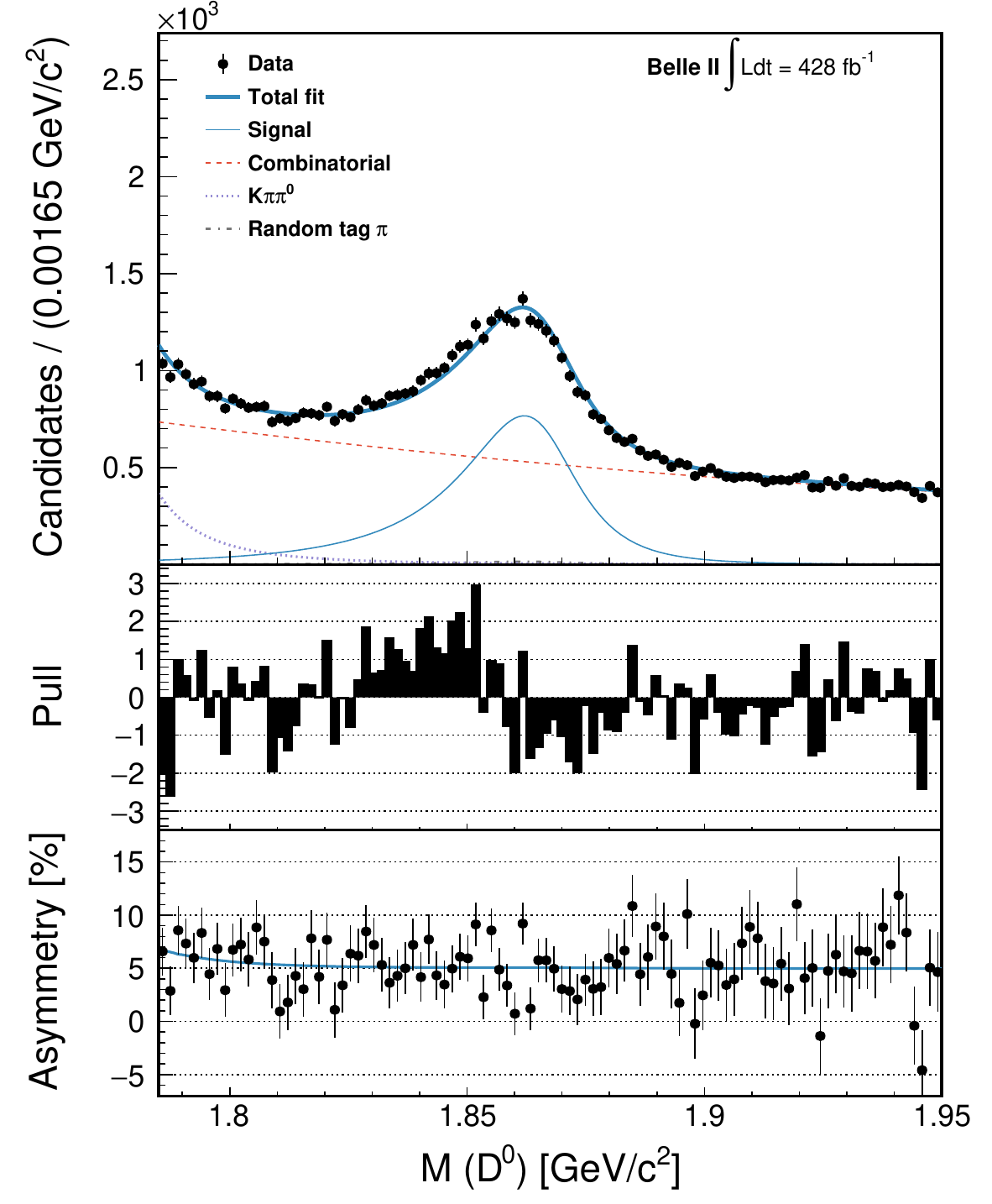}
    \includegraphics[width=0.49\linewidth]{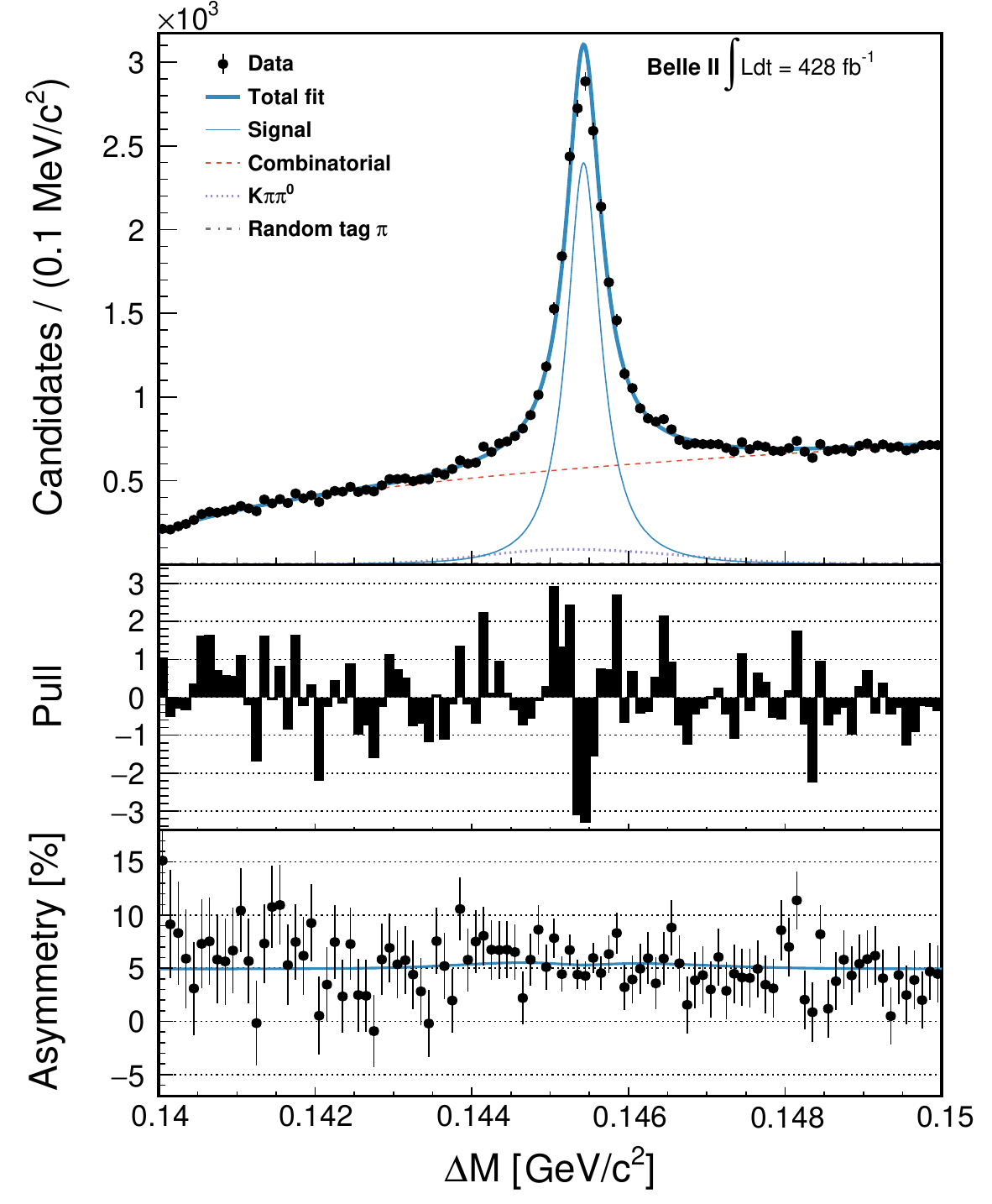}
    \caption{\DzTopipipiz sample: distributions of the \Dz candidate invariant mass (left) and \dm (right) for \(\Dzcosthetacm\in[-0.208,0)\), with fit functions overlaid. The middle plots of each panel show the pull (difference between data and fit result divided by the data uncertainty). The bottom plots show the \Dz--\Dzb asymmetry of the data (black points) and of the total PDF (blue line), computed for each bin using \autoref{eq:araw}.\label{fig:datafit}}
\end{figure*}

\begin{figure}[b]
    \includegraphics[page=1,width=\linewidth]{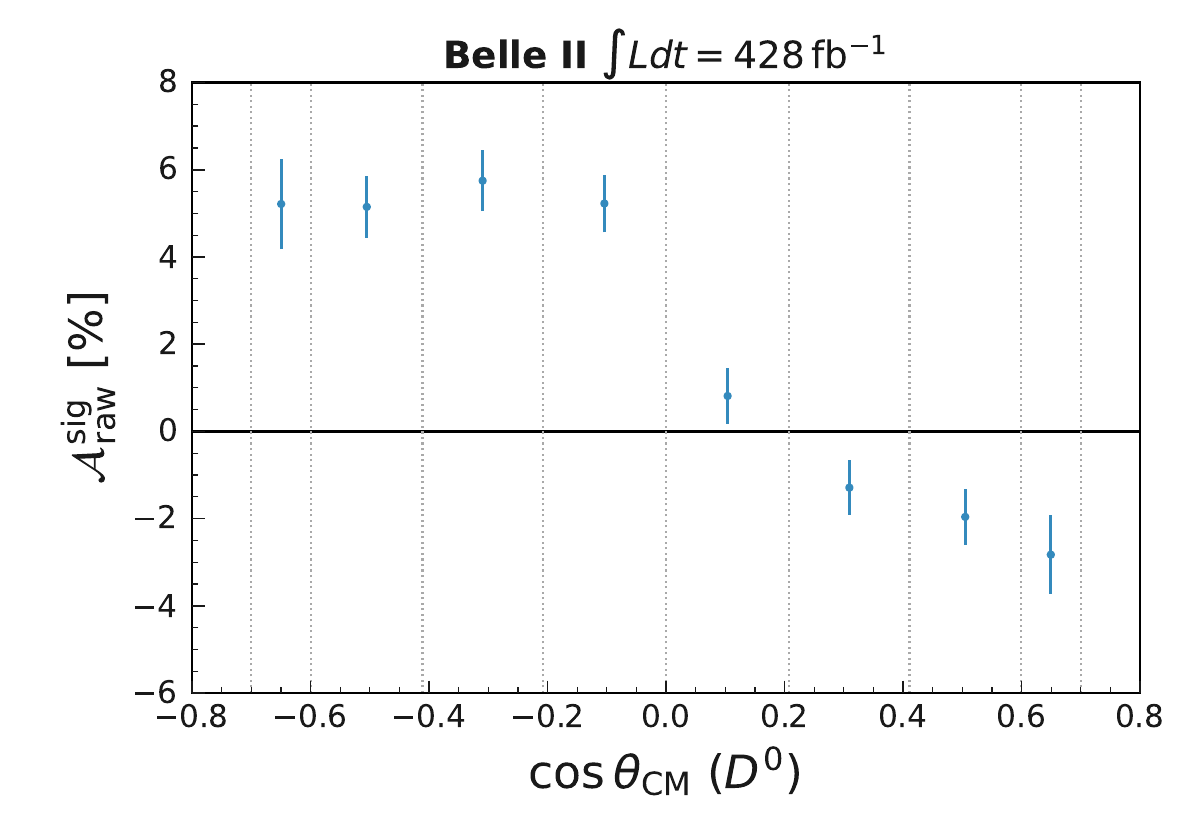}
    \caption{\DzTopipipiz raw asymmetries from the fit in the eight \Dzcosthetacm bins. The error bars show statistical uncertainties.\label{fig:dataRawA}}
\end{figure}

\autoref{fig:datafit} shows the \(M\) and \dm distributions, for one \Dzcosthetacm bin, with fit projections overlaid.
Some residual mismodeling is evident from the data-fit pulls;
while this may seem a large effect at first, the sample asymmetry as a function of the fit variables is well reproduced by the fit function, and the impact on the measurement is very small, as discussed in more detail in \autoref{sec:syst}.
The signal yield integrated over all \Dzcosthetacm bins is \((271.4\pm0.7)\times10^3\).
\autoref{fig:dataRawA} shows the resulting raw asymmetries as a function of \Dzcosthetacm.

\subsection{Tagged \DzToKpiRS sample\label{ss:fits:tagged}}
We use the \dm distribution to measure the tagged sample asymmetries.
We consider only two components: correctly reconstructed and tagged \DzToKpiRS decays, and a background consisting of both correctly reconstructed \DzToKpiRS decays associated to a random tag pion and random combinations of final-state particles.
Background due to misreconstructed charm decays, \eg, \(\Dz\to\Km\mup\neumb\) where the muon is misidentified as a pion and the neutrino is not reconstructed, is negligible.

The \DzToKpiRS decay distribution is modeled by the sum of one Johnson's \(S_U\) and two Gaussian distributions, with one shared location parameter.
Width parameters are different for \Dz and \Dzb candidates to account for differing momentum scales and resolutions for positive and negative final-state particles.
This happens because, due to the detector geometry, positive and negative particle trajectories intersect different detector regions and amounts of material.
The background is modeled with a threshold-like function, as in \autoref{eq:pdf:thr}.

\begin{figure}[htbp]
    \adjincludegraphics[page=1,width=\linewidth,trim={{0.4\width} 0 {0.4\width} 0},clip]{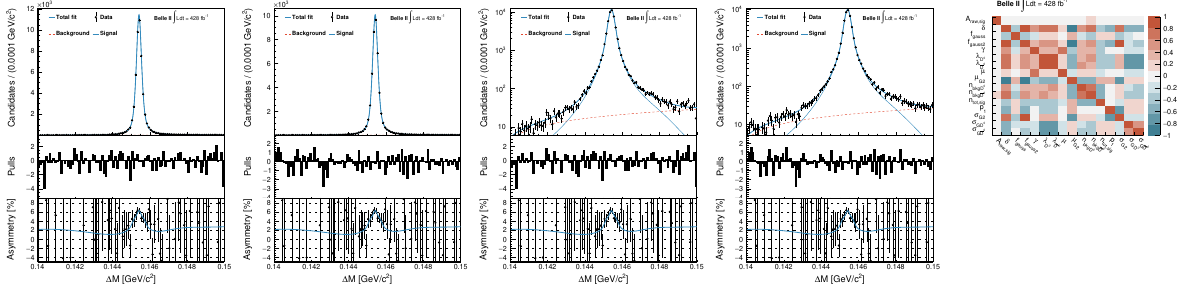}
    \caption{\DzToKpiRS tagged sample: \dm distribution with fit functions overlaid, for \(\Dzcosthetacm\in[-0.208,0)\). The middle plot shows the pull, while the bottom plot shows the \Dz--\Dzb asymmetry of the data (black points) and the total PDF (blue line), computed for each bin using \autoref{eq:araw}.\label{fig:fits:tagged}}
\end{figure}

Each \Dzcosthetacm bin is fitted independently.
\autoref{fig:fits:tagged} shows the \dm distribution of the data, for one \Dzcosthetacm bin, with fit projections overlaid.
The structure in the asymmetry plot is produced by the aforementioned different \dm resolution for \Dz and \Dzb mesons.
The \DzToKpiRS yield integrated over all \Dzcosthetacm bins is \((744.4\pm1.1)\times10^3\).

\subsection{Untagged \DzToKpiRS sample\label{ss:fits:untagged}}
We use the \(M\) distribution to measure the untagged sample asymmetry.
We consider only two components: correctly reconstructed \DzToKpiRS decays, and background from all other sources that do not produce a peak in \(M\).
Peaking background from ``wrong-sign'' \DzToKpiWS decays, which arise from doubly Cabibbo-suppressed \DzToKpiWS decays and Cabibbo-favored \DzbToKpiRS decays preceded by \Dz--\Dzb mixing, are neglected. They amount to \(\sim0.4\%\) of the \DzToKpiRS yield and dilute the measured asymmetry, since they provide the incorrect flavor tag.
A systematic uncertainty is assigned for this in \autoref{sec:syst}.

The \DzToKpiRS decay distribution is modeled by the sum of a Johnson and a Gaussian distribution, with one shared location parameter.
Width parameters are different for the two flavors as for the tagged sample.
The background is modeled with a first-order polynomial.

Each \Dzcosthetacm bin is fitted independently.
\autoref{fig:fits:untagged} shows the \(M\) distribution, for one \Dzcosthetacm bin, with fit projections overlaid.
The \DzToKpiRS yield integrated over all \Dzcosthetacm bins is \((3232\pm4)\times10^3\).

\begin{figure}[htbp]
    \adjincludegraphics[page=2,width=\linewidth,trim={0 0 {0.8\width} 0},clip]{data-sample-fits}
    \caption{\DzToKpiRS untagged sample: \(M\) distribution with fit functions overlaid, for \(\Dzcosthetacm\in[-0.208,0)\). The middle plot shows the pull, while the bottom plot shows the \Dz--\Dzb asymmetry of the data (black points) and the total PDF (blue line), computed for each bin using \autoref{eq:araw}.\label{fig:fits:untagged}}
\end{figure}

\begin{figure}[htbp]
    \includegraphics[page=8,width=\linewidth]{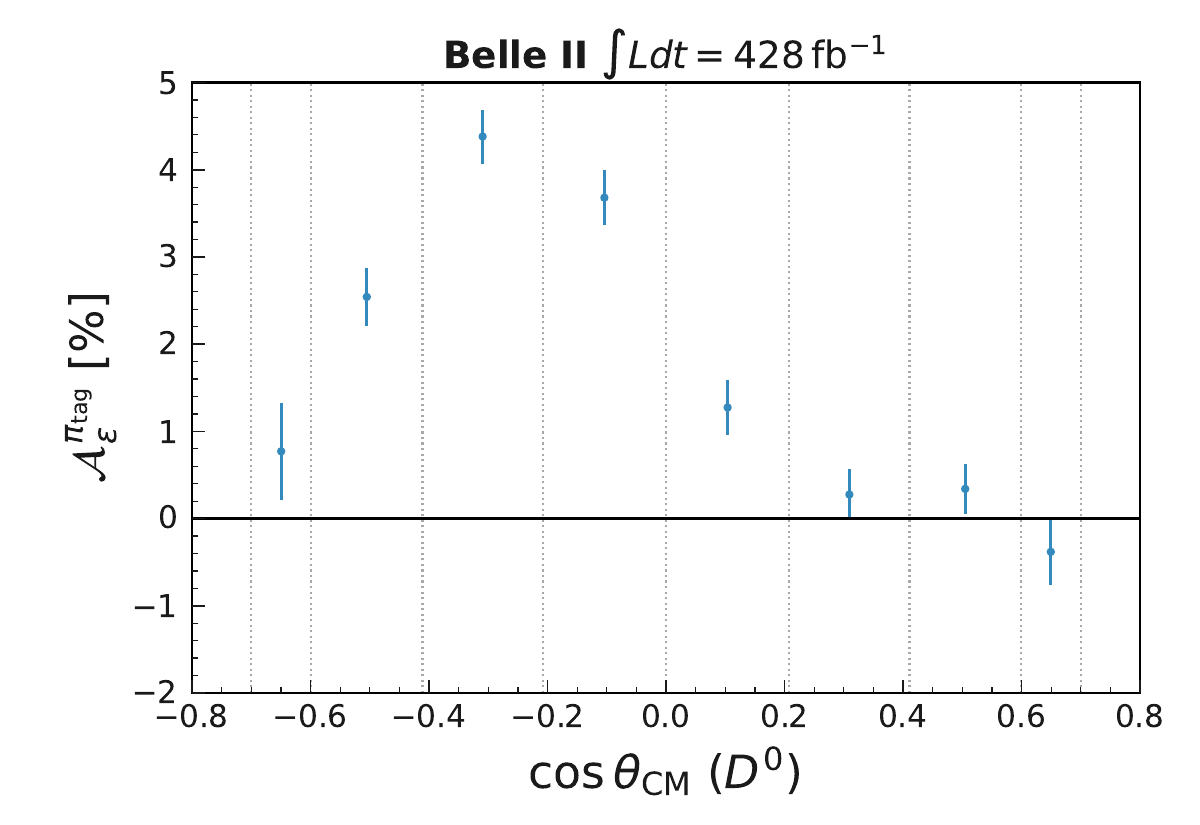}
    \caption{\Adetpitag computed from the \DzToKpiRS tagged and untagged raw asymmetries using \autoref{eq:AtagDef}, in the eight \Dzcosthetacm bins. The error bars show statistical uncertainties.\label{fig:fits:Atag}}
\end{figure}

\autoref{fig:fits:Atag} shows the tag pion reconstruction asymmetry computed using \autoref{eq:AtagDef}.
As the tagged decays are a subset of the untagged decays, we take into account the correlation between the two control samples when computing the uncertainty on \Adetpitag.

\subsection{\texorpdfstring{\(\ACP(\DzTopipipiz)\)}{ACP(D0 -> pi+ pi- pi0)} determination}
\begin{figure}[htbp]
    \includegraphics[page=3,width=\linewidth]{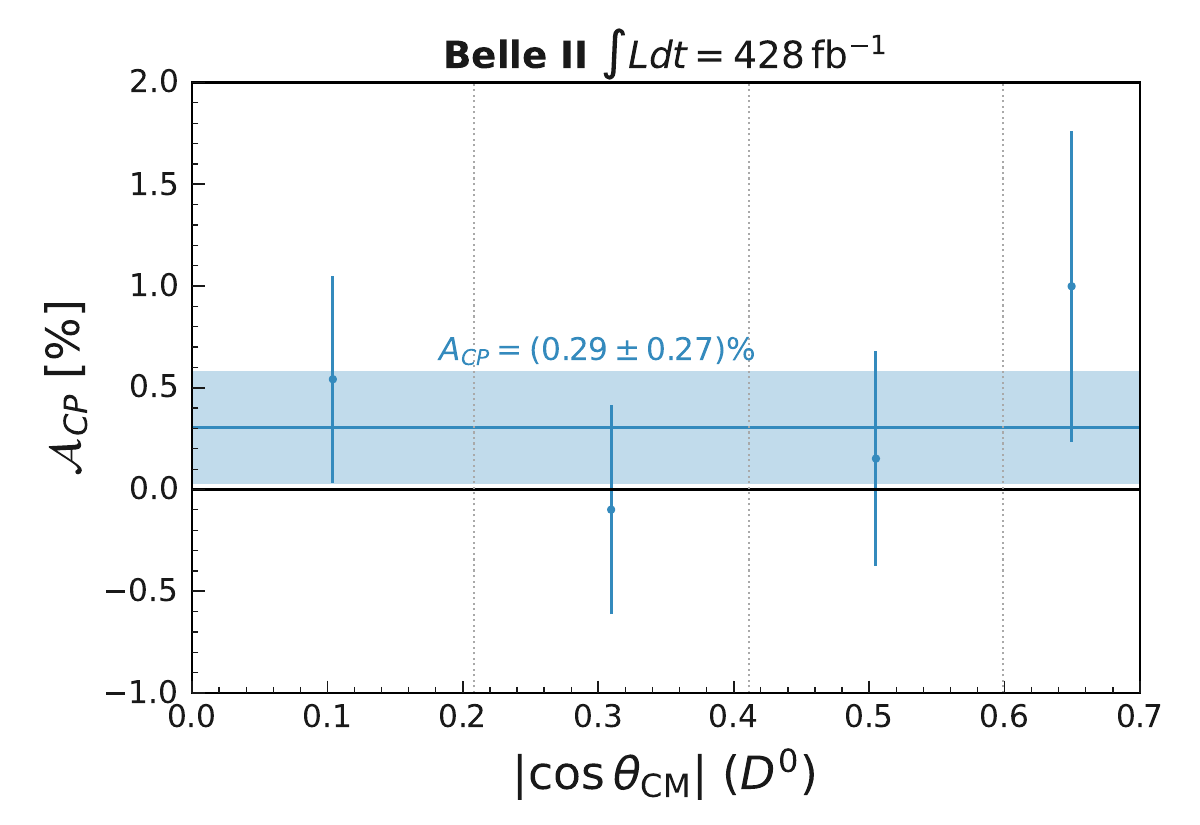}
    \caption{\(\ACP^i\) in the four \Dzcosthetacm bin pairs, and the average \ACP value. The uncertainties are statistical and include contributions from both the signal and control channels.\label{fig:fits:Afinal}}
\end{figure}

\autoref{fig:fits:Afinal} shows the \(\ACP^i\) determinations in the four \Dzcosthetacm bin pairs, computed using \autoref{eq:AfinalDef}.
Averaging these measurements, we obtain
\begin{equation}
    \ACP(\DzTopipipiz) = (0.29\pm0.27)\%\,,
\end{equation}
where the uncertainty is statistical and includes the uncertainty from \Adetpitag (\(0.12\%\)).

\section{Systematic uncertainties\label{sec:syst}}
We consider six sources of systematic uncertainty:
the \Dz reconstruction asymmetry;
the fact that some \DzTopipipiz fit parameters are fixed to the values determined from simulation;
the biases of the fits;
the residual mismodeling of the PDFs;
the uncertainty arising from the control sample weighting procedure;
and the presence of wrong-sign \DzToKpiWS decays in the \DzToKpiRS control samples.
\autoref{tab:syst} summarizes their impact.

\begin{table}[hbp]
    \caption{Summary of the systematic uncertainties affecting the \ACP measurement. The total systematic uncertainty is the quadrature sum of the individual components. The statistical uncertainty is reported for reference.\label{tab:syst}}
    \begin{ruledtabular}
        \begin{tabular}{lc}
            Source & Uncertainty \\
            \colrule
            Signal channel \\
            \quad \Dz reconstruction asymmetry & \(0.06\%\) \\
            \quad Fit parameters fixed from MC & \(0.02\%\) \\
            \quad Fit bias & \(0.06\%\) \\
            \quad PDF mismodeling & \(0.03\%\) \\
            Control channel \\
            \quad Fit bias & \(0.04\%\) \\
            \quad PDF mismodeling & \(0.07\%\) \\
            \quad Weighting & \(0.03\%\) \\
            \quad Wrong-sign \DzToKpiWS decays & \(0.01\%\) \\
            \textbf{Total systematic} & \(\mathbf{0.13\%}\) \\
            \colrule
            \textbf{Total statistical} & \(\mathbf{0.27\%}\) \\
        \end{tabular}
    \end{ruledtabular}
\end{table}

We estimate the \Dz reconstruction asymmetry using the MC simulation to be \((8\pm6)\times10^{-4}\), which is consistent with zero as expected.
Since the \DzTopipipiz Dalitz distribution model is based on the decay amplitude model measured by the \babar Collaboration \cite{BaBarAmplitude}, we expect the MC to accurately reproduce the kinematic distributions of the \Dz pions.
The simulation shows that the \Dz pions typically have large transverse momenta.
Dedicated studies determined the data-simulation agreement for charge asymmetries to be good in this region of phase space; therefore, we can rely on the value obtained from MC.
We assign the statistical uncertainty of this determination as the associated systematic uncertainty.

The uncertainty due to fixed PDF parameters for the signal channel fit is evaluated by varying these parameters by their uncertainties and repeating the fit.
We sample all parameters simultaneously from Gaussian distributions having widths equal to their uncertainties and repeat this sampling and subsequent fitting 2000 times.
During this sampling, we account for correlations among the parameters.
We record the 2000 fitted values of \ACP and assign the standard deviation of their distribution as the related systematic uncertainty.
The impact of fixing the random-tag-pion fraction, misreconstructed signal fraction, and random-tag-pion asymmetry is evaluated separately using the same procedure and repeating the fit 100 times.
To account for possible differences between data and simulation, we conservatively use \(25\%\) of the fractions' nominal values as the standard deviations.
This corresponds to shifts of \(\sim0.5\%\) for the random-tag-pion fraction and \(\sim1.3\%\) for the fraction of misreconstructed signal.
The sum in quadrature of the standard deviations of the \ACP values, \((1.89\pm0.06)\times10^{-4}\), is assigned as the systematic uncertainty.
The dominant contributions to this uncertainty are those from the misreconstructed signal fraction and from the fixed PDF parameters.

We evaluate possible fit biases using linearity tests based on pseudoexperiments, which are generated from the PDFs fitted to data with different input values of the signal raw asymmetry.
We then compare the fitted asymmetry to the input value and
obtain average differences of \((-6.2\pm0.6)\times10^{-4}\) for the signal sample, \((2.2\pm0.4)\times10^{-4}\) for the tagged \DzToKpiRS sample, and \((-3.2\pm0.2)\times10^{-4}\) for the untagged \DzToKpiRS sample.
We assign the absolute values of these biases as systematic uncertainties, which are added in quadrature.

We evaluate the impact of the residual mismodeling of the PDFs as follows.
Most of the mismodeling is due to the assumption that \(M\) and \dm PDFs factorize into one-dimensional PDFs, which neglects some existing correlations for some of the components (especially \DzToKpipiz).
The mismodeling seen in data is well reproduced by the simulation.
We evaluate a systematic uncertainty using bootstrap subsamples \cite{bootstrap}, \ie, subsamples obtained by randomly drawing candidates from the MC sample, with replacement.
The bootstrap subsamples have the same size as the data.
We repeat the fit on 1000 bootstrap subsamples, compute the difference between the fitted asymmetry and its true value, and use the mean of the distribution of differences as the bias estimate.
For the signal sample, we repeat this procedure five times using
the unmodified MC sample, where \ACP is zero;
a weighted MC sample where the signal raw asymmetry is increased by a factor of \(2\);
a weighted MC sample where the signal raw asymmetry is reduced by a factor of \(2\);
a weighted MC sample where the true \ACP is increased to \(+1\%\); and
a weighted MC sample where the true \ACP is reduced to \(-1\%\).
The maximum bias we obtain is \((2.7\pm0.7)\times10^{-4}\).
For the control samples, we bootstrap tagged and untagged samples together in order to preserve their correlation and obtain \((-7.1\pm0.4)\times10^{-4}\).
We assign the absolute values of these biases as systematic uncertainties.

We evaluate the impact of the kinematic weighting of the control samples by repeating the \Adetpitag measurement without weights.
The difference in \ACP is \((-5.8\pm1.0)\times10^{-4}\).
We conservatively assign half of the magnitude of this difference as the systematic uncertainty, covering effects both from \splt weights and final sample weights.

We evaluate the impact of the wrong-sign \DzToKpiWS decays by computing the dilution of the untagged sample asymmetry due to their fraction, assuming a \(100\%\) flavor mistag rate.
We obtain \((7.56\pm0.03)\times10^{-5}\).

We perform several consistency checks.
We repeat the measurement on data with five different \Dzcosthetacm binnings and find that the \ACP measurements are all compatible within the statistical uncertainty (taking into account that the samples are fully correlated).
We test alternative fit models for the signal channel (different parameter splitting among \Dzcosthetacm bins, parameter splitting by flavor, alternative PDFs), and find no statistically significant impact on the \ACP measurement.
Finally, we repeat the measurement after dividing the data sample into four bins of \Dz azimuthal angle and five data-taking periods.
We find all results to be compatible with each other and with the nominal result.

\section{Conclusions\label{sec:results}}
Using \Dstarp-tagged \DzTopipipiz decays reconstructed in the \bii dataset collected between 2019 and 2022, which corresponds to an integrated luminosity of \lumi, we measure the time-integrated \CP asymmetry in \DzTopipipiz decays to be
\begin{equation}
    \ACP = (0.29\pm0.27\pm0.13)\%\,,
\end{equation}
where the first uncertainty is statistical and includes the contribution from the \DzToKpiRS control samples used to correct experimentally-induced asymmetries and the second uncertainty is systematic.
The result is consistent with \CP symmetry and with existing measurements \cite{babarACP,belleACP}.
The result is \(34\%\) more precise than the current world's best measurement from \babar \cite{babarACP}, despite an increase of only about \(10\%\) in integrated luminosity.
The increase in precision per unit luminosity can be attributed to the novel candidate selection and analysis strategy employed.

\section*{Acknowledgments}
This work, based on data collected using the Belle II detector, which was built and commissioned prior to March 2019,
was supported by
Higher Education and Science Committee of the Republic of Armenia Grant No.~23LCG-1C011;
Australian Research Council and Research Grants
No.~DP200101792, 
No.~DP210101900, 
No.~DP210102831, 
No.~DE220100462, 
No.~LE210100098, 
and
No.~LE230100085; 
Austrian Federal Ministry of Education, Science and Research,
Austrian Science Fund (FWF) Grants
DOI:~10.55776/P34529,
DOI:~10.55776/J4731,
DOI:~10.55776/J4625,
DOI:~10.55776/M3153,
and
DOI:~10.55776/PAT1836324,
and
Horizon 2020 ERC Starting Grant No.~947006 ``InterLeptons'';
Natural Sciences and Engineering Research Council of Canada, Digital Research Alliance of Canada, and Canada Foundation for Innovation;
National Key R\&D Program of China under Contract No.~2024YFA1610503,
and
No.~2024YFA1610504
National Natural Science Foundation of China and Research Grants
No.~11575017,
No.~11761141009,
No.~11705209,
No.~11975076,
No.~12135005,
No.~12150004,
No.~12161141008,
No.~12405099,
No.~12475093,
and
No.~12175041,
and Shandong Provincial Natural Science Foundation Project~ZR2022JQ02;
the Czech Science Foundation Grant No. 22-18469S,  Regional funds of EU/MEYS: OPJAK
FORTE CZ.02.01.01/00/22\_008/0004632 
and
Charles University Grant Agency project No. 246122;
European Research Council, Seventh Framework PIEF-GA-2013-622527,
Horizon 2020 ERC-Advanced Grants No.~267104 and No.~884719,
Horizon 2020 ERC-Consolidator Grant No.~819127,
Horizon 2020 Marie Sklodowska-Curie Grant Agreement No.~700525 ``NIOBE''
and
No.~101026516,
and
Horizon 2020 Marie Sklodowska-Curie RISE project JENNIFER2 Grant Agreement No.~822070 (European grants);
L'Institut National de Physique Nucl\'{e}aire et de Physique des Particules (IN2P3) du CNRS
and
L'Agence Nationale de la Recherche (ANR) under Grant No.~ANR-23-CE31-0018 (France);
BMFTR, DFG, HGF, MPG, and AvH Foundation (Germany);
Department of Atomic Energy under Project Identification No.~RTI 4002,
Department of Science and Technology,
and
UPES SEED funding programs
No.~UPES/R\&D-SEED-INFRA/17052023/01 and
No.~UPES/R\&D-SOE/20062022/06 (India);
Israel Science Foundation Grant No.~2476/17,
U.S.-Israel Binational Science Foundation Grant No.~2016113, and
Israel Ministry of Science Grant No.~3-16543;
Istituto Nazionale di Fisica Nucleare and the Research Grants BELLE2,
and
the ICSC – Centro Nazionale di Ricerca in High Performance Computing, Big Data and Quantum Computing, funded by European Union – NextGenerationEU;
Japan Society for the Promotion of Science, Grant-in-Aid for Scientific Research Grants
No.~16H03968,
No.~16H03993,
No.~16H06492,
No.~16K05323,
No.~17H01133,
No.~17H05405,
No.~18K03621,
No.~18H03710,
No.~18H05226,
No.~19H00682, 
No.~20H05850,
No.~20H05858,
No.~22H00144,
No.~22K14056,
No.~22K21347,
No.~23H05433,
No.~26220706,
and
No.~26400255,
and
the Ministry of Education, Culture, Sports, Science, and Technology (MEXT) of Japan;  
National Research Foundation (NRF) of Korea Grants
No.~2021R1-F1A-1064008, 
No.~2022R1-A2C-1003993,
No.~2022R1-A2C-1092335,
No.~RS-2016-NR017151,
No.~RS-2018-NR031074,
No.~RS-2021-NR060129,
No.~RS-2023-00208693,
No.~RS-2024-00354342
and
No.~RS-2025-02219521,
Radiation Science Research Institute,
Foreign Large-Size Research Facility Application Supporting project,
the Global Science Experimental Data Hub Center, the Korea Institute of Science and
Technology Information (K25L2M2C3 ) 
and
KREONET/GLORIAD;
Universiti Malaya RU grant, Akademi Sains Malaysia, and Ministry of Education Malaysia;
Frontiers of Science Program Contracts
No.~FOINS-296,
No.~CB-221329,
No.~CB-236394,
No.~CB-254409,
and
No.~CB-180023, and SEP-CINVESTAV Research Grant No.~237 (Mexico);
the Polish Ministry of Science and Higher Education and the National Science Center;
the Ministry of Science and Higher Education of the Russian Federation
and
the HSE University Basic Research Program, Moscow;
University of Tabuk Research Grants
No.~S-0256-1438 and No.~S-0280-1439 (Saudi Arabia), and
Researchers Supporting Project number (RSPD2025R873), King Saud University, Riyadh,
Saudi Arabia;
Slovenian Research Agency and Research Grants
No.~J1-50010
and
No.~P1-0135;
Ikerbasque, Basque Foundation for Science,
State Agency for Research of the Spanish Ministry of Science and Innovation through Grant No. PID2022-136510NB-C33, Spain,
Agencia Estatal de Investigacion, Spain
Grant No.~RYC2020-029875-I
and
Generalitat Valenciana, Spain
Grant No.~CIDEGENT/2018/020;
The Knut and Alice Wallenberg Foundation (Sweden), Contracts No.~2021.0174, No.~2021.0299, and No.~2023.0315;
National Science and Technology Council,
and
Ministry of Education (Taiwan);
Thailand Center of Excellence in Physics;
TUBITAK ULAKBIM (Turkey);
National Research Foundation of Ukraine, Project No.~2020.02/0257,
and
Ministry of Education and Science of Ukraine;
the U.S. National Science Foundation and Research Grants
No.~PHY-1913789 
and
No.~PHY-2111604, 
and the U.S. Department of Energy and Research Awards
No.~DE-AC06-76RLO1830, 
No.~DE-SC0007983, 
No.~DE-SC0009824, 
No.~DE-SC0009973, 
No.~DE-SC0010007, 
No.~DE-SC0010073, 
No.~DE-SC0010118, 
No.~DE-SC0010504, 
No.~DE-SC0011784, 
No.~DE-SC0012704, 
No.~DE-SC0019230, 
No.~DE-SC0021274, 
No.~DE-SC0021616, 
No.~DE-SC0022350, 
No.~DE-SC0023470; 
and
the Vietnam Academy of Science and Technology (VAST) under Grants
No.~NVCC.05.02/25-25
and
No.~DL0000.05/26-27.

These acknowledgements are not to be interpreted as an endorsement of any statement made
by any of our institutes, funding agencies, governments, or their representatives.

We thank the SuperKEKB team for delivering high-luminosity collisions;
the KEK cryogenics group for the efficient operation of the detector solenoid magnet and IBBelle on site;
the KEK Computer Research Center for on-site computing support; the NII for SINET6 network support;
and the raw-data centers hosted by BNL, DESY, GridKa, IN2P3, INFN, 
and the University of Victoria.

\section*{Data Availability}
The full \bii data are not publicly available.
The collaboration will consider requests for access to the data that support this article.

\bibliography{references}

\end{document}